\documentclass[12pt,a4paper]{article}
\usepackage[T1]{fontenc}
\usepackage{calc}
\usepackage{hyperref}\usepackage{setspace}
\usepackage{multicol}
\usepackage[normalem]{ulem}
\usepackage{color}
 
\begin{document}

\begin{center}
{\large \bf Massless Electron and Fractional Spin as Electronic Charge } \\
{\bf S.C. Tiwari} \\
{\it Institute of Natural Philosophy} \\ 
{\it 1, Kusum Kutir, Mahamanapuri, Varanasi - 221 005, India} \\
{\it email}: vns\_sctiwari@yahoo.com
\end{center}
\begin{center}
{\bf Abstract}
\end{center}

The standard model (SM) of particle physics has been supported by several
experimental findings, the most remarkable of them being the discovery of the
weak gauge bosons , W and Z. It is expected that the Higgs boson could show up
by 2007 at LHC, CERN. In spite of this, the unsatisfactory features of the SM
at conceptual level, and exclusion of gravity from the unification scheme have
led to explore 'the physics beyond the SM'.

A critique and comprehensive review of the contemporary fundamental physics
was presented in a monograph completed in the centenary year,1997 of the
discovery of the electron. A radically new approach to address foundational
problems was outlined: masslessness of bare electron ,interpretation of the
squared electronic charge in terms of the fractional spin, $e^2/c$; new physical
significance of the electromagnetic potentials, 2+1 dimensional internal
structure of electron and neutrino, and composite photon are some of the ideas
proposed. Though the monograph was reviewed by E. J. Post(Physics
Essays, June1999), it has remained largely inaccessible. I believe some of
these unconventional ideas have a potential to throw light on the fundamental
questions in physics, and therefore deserve a wider dissemination.

The reader may find illuminating to supplement Sec. 3 on the weak gauge 
bosons with a candid,graceful and personal recollection by 
Pierre Darriulat(CERN Courier, April 2004, p.13).

\section{Introduction}

Reality is simple and comprehensible to the human mind. Or is it
complex  and beyond intellect? It would seem that if it is simple then it is 
not interesting, and the claim of the perception of the reality would be viewed 
with disbelief. Paradoxically, however, it is the search of the ultimate
reality whether in the form of the elementary constituents of the matter or a
unified theory of nature which has been the prime goal of science. Is this
goal attainable? I think a scientist can glimpse the reality as a personal
subjective experience, however, a fundamental scientific theory for the
ultimate reality or its representation in a physical form will never be
possible. Quite often, the most intense and fruitful debate in science has
occurred due to the conflicting beliefs of the original minds. Certitude in
their beliefs may have some relationship with their transcendental experiences
to perceive the reality.
Pathways to such experiences influence one's mind so strongly that one is led
to view everything within that framework. Doubts and reexamination of such
beliefs dispassionately are necessary to free them from the transcendental  
elements. Rarely do we find also the liberated minds who are not attached to
their ideas. The writings of Henry Poincare give an impression of his being
such a liberated thinker \cite{1}. It may be pointed out that detachment
should not be viewed as sterile and inactive mind; a passionate enquirer may
afford to be a liberated mind. Excessive attachment may hamper creativity and
originality of an individual, and may harm scientific discourses on the
fundamental problems. The role of unobservables and metaphysical elements in a
physical theory seem to be related; the Bohr-Einstein debate on the 
interpretation and foundations of quantum mechanics, and the nature of time
are discussed in \cite{2} from this viewpoint. In this monograph I will adopt
thus outlook both for the criticisms of the established theories and for
alternative propositions. 

The subject matter of this study, the electron, discovered a hundred years ago,
may appear obsolete, and give the impression of moving backward in time. In
the light of the vast knowledge accumulated since the discovery of the
electron, announced on 29 April, 1897 in a Friday Evening Discourse at the
Royal Institution by J.J. Thomson, my aim is to analyze the fundamental
problems related with the electron in a modern perspective, and to propound 
a radically new alternative approach. This book is, therefore, not a historical
treatise or a popular account of the discovery of the electron. It is worth
explaining the necessity and the significance of this work.

In the Review of Particle Physics \cite{3}, the electron is listed as just one
amongst hundreds of elementary particles, and precise measured values of its
physical properties are given. So, what else is there to discuss? A
discomforting feature is immediately obvious: the periodic table of the
elements has 112 entries, while the number of the elementary particles is very
large as compared to this. Not only this, the much acclaimed unified gauge
theory does not explain the physical meaning of the elementary properties like
charge, mass, and spin of, let us say the electron, while at the same time
many more new quantum numbers are introduced. A brief  	non-technical
discussion on the elementary particle physics may be helpful to appreciate
these remarks. 

From the extract of M. Ampere's proposition given in the Source Book by Magie
(M. p.114) we learn that the name particle is given to an infinitely small
portion of a body of the same nature as the body; the particles are made up of
molecules, and a molecule is an assemblage of atoms. In contrast to this the
modern treatment postulates elementary particles as the constituents of atoms.
Initially the idea that atoms are composite structures of elementary particles
was indeed quite attractive as there were very few such constituent particles,
e.g. electron, proton and neutron. As for the interaction between them,
amongst the four fundamental forces of nature, the gravitational force and the
electromagnetic force have been known since long, and both are long range
forces. 
The gravitational force is so weak that it does not seem to be important in
elementary particle interactions, to be precise at the practical energy
scales. The other two forces are short range forces with highly different
strengths, termed as weak and strong forces. The advent of quantum theory led
to a new language and description for the forces: field quanta and their
exchange during interactions. For quite some time there were only the
electromagnetic field quantum: photon, and Yukawa's hypothesized strong field
quanta : mesons. Binding of nucleons (neutron and proton) in the nuclei of
atoms is explained by postulating strong force, and the beta-decay by the weak
force. Later discoveries of new particles necessitated the classification of
them as leptons which are not sensitive to the strong force, and rest of them
as hadrons. The scheme that atoms and nuclei are composite objects of the
elementary building blocks i.e. neutron, proton and electron was extended to
the newly discovered hadrons. Fermi and Yang suggested pion ($\pi$-meson) as a
bound nucleon-antinucleon state. Sakata accounted for strange hadrons in terms
of the triplet of proton, neutron and lambda particles with the symmetry group
SU(3). Baryon is a hadron obeying the Fermi-Dirac statistics i.e. a fermion,
and meson is a hadron obeying the Bose-Einstein statistics i.e. a boson.

G. Zweig and M. Gell-Mann in 1964 independently proposed the quark model for
hadrons. In the earliest version of the triplet model, three types of basic
building blocks called quarks are postulated possessing spin half, and
fractional electric charge and baryon numbers The mesons are quark-antiquark
bound states, and baryons have three quarks as their constituents. Now we have
quarks characterized by six flavors, and each flavor comes in three colors.
The color is strong interaction charge, analogous to the electric charge in
electromagnetic interactions. There is a difference, however, the strong field
quanta called gluons carry color unlike the photon which is neutral. The
hadrons are postulated to be colorless composite particles. Even the quarks
and leptons as elementary particles are not small in number. Speculating
sub-leptons and sub-quarks exotic models also exist in the literature. The
question arises: Is this philosophy satisfactory? I do not think in this way 
we may know the ultimate constituents of the matter. An alternative view-point
could be that the space-time structure itself is the most elementary entity,
and the matter is a manifestation of the geometry of the space-time. This idea
makes physical sense if there is a viable scheme for the elementary particle
model. Amongst the known elementary particles the stable ones are electron
(positron), neutrinos, photon and proton. Excluding the heavy particle proton,
it is possible that massless particles and electron could be the elementary
constituents of matter. 
To explore this idea we need to understand them at a basic level, and direct
our attention on the electron because it is distinct from neutrinos and photon
since it has charge and mass.

Related to the elementary constituents, there is the question of fundamental
interactions and their unification. The interaction between the quarks and the
leptons is believed to be described by the Standard Model (SM) which is a
gauge theory of strong, weak and electromagnetic interactions. The principle
of gauge symmetry, and formal structure of quantum electrodynamics have played
pivotal role in the development of the SM. The gauge group in QED is U(1)
while the SM has the gauge group $SU(3) \times SU(2) \times U(1)$. The gauge theory of
strong interaction in color space is called quantum chromodynamics  (QCD) with
the gauge group $SU(3)$. The effective coupling constant in QCD is energy
dependent decreasing with increasing energy. This phenomenon is known as
asymptotic freedom. 
Unified theory of weak and electromagnetic interactions i.e. the electroweak
theory is described by the gauge group $SU(2) \times U(1)$. Besides photon, there are
weak vector gauge bosons, $W^{\mathrm{+}}$, $W^{\mathrm{\-}}$ and $Z$ in this
theory. Both QCD and the electroweak theory have been proved to be
renormalisable theories. Wider symmetry groups like $SU(5)$ for grand
unification in view of the opposite energy dependence of the coupling
constants of QCD and electroweak theory have been proposed. Supersymmetric
generalizations of the SM using the supersymmetry which relates bosons and
fermions have also been made. Discovery of neutral weak current, weak gauge
bosons and quarks signatures in high energy collisions, and precision
measurements in particle physics give confidence in the SM. Weaknesses of the
SM have also been noted by the physicists, and incorporation of gravitational
force is considered as one of the most outstanding problems in the unification
scheme. In the next chapter, we will discuss some of the questions related
with the Standard Model, here we point out the most unsatisfactory aspect of
the SM: it does not explain the meaning of charge and mass (of the electron).
The electroweak theory is hailed as a great synthesis in the spirit of the
Maxwell theory of electromagnetism. Unfortunately, none of the foundational  	
problems are solved: structure of the electron, duality of the source and the
field, incurable infinities in point field theory, the meaning of charge, mass
and spin. 

Postulating unobservable particles like quarks and gluons, and introducing many
more quantum numbers or charges in abstract internal spaces in a theory aimed
at unification indicate fundamental flaw in the approach. Most of the
elementary particles are inferred from the expected indirect signatures (like
decay modes). T.D. Lee has remarked \cite{4} that ``The progress of particle
physics is closely tied to the discovery of resonances, which started at the  
Chicago Cyclotron. Yet even the great Enrico Fermi, when he proposed the
machine, did not envisage this at all. After the discovery of the first
nucleon resonance, for almost a year Fermi expressed doubts whether it was
genuine''.  Proliferation of new particles in the laboratory, and the remote
methods of their observations which include arbitrariness in the fitting
parameters at least call for a cautious approach in attributing them the
physical reality.  The words of William Crookes \cite{5} are probably quite
relevant today : ``I hope I may be allowed to record some theoretical
speculations which have gradually formed in my mind during the progress of
these experiments. I put them forward only as working hypotheses, useful,
perhaps necessary, in first dawn of new knowledge, but only to be retained as
long as they are of assistance; for experimental research is necessarily and
slowly progressive, and one's early provisional hypotheses have to be
modified, adjusted, perhaps altogether abandoned in deference to later
observations''. It is true that the quark model which is now one of the basic
ingredients of the Standard Model, was not acceptable to the physicists when
it was first proposed. Kendall mentions the story how Zweig's paper could not
be published until the mid 1970s \cite{6}. But the current dogmatic faith in
the SM is also unwarranted; the dissatisfaction with the trends in particle
physics is best summed up by Dirac \cite{7}: ``Still, that was the situation  
in those days; people were very reluctant to postulate a new particle. The
situation is quite different nowadays, when people are only too willing to
postulate a new particle on the slightest evidence, either theoretical or  
experimental''.

Experience in particle physics so far has unambiguously shown that
understanding the ultimate constituents of matter by performing high energy
scattering experiments and the current paradigm for developing a ``theory of 
everything'' are proving to be mirages. There is, therefore, a need to look
for an alternative approach to deal with the fundamental problems in physics. I
think understanding the electron and the electromagnetism may throw light in 
this direction. We know that the gauge theories in various forms have
underlying guiding theory that of the electromagnetic fields, and the
development of quantum theory was inspired by the ``radical revision of
classical dynamics for the electron''. The existing reviews or discussions
adopt the viewpoint dividing the problems at classical or quantum level. But
the theories are the descriptions of the phenomena, not the phenomena,
therefore, this separation is artificial obscuring the physical origin of the
problems. In this monograph, an integrated and constructive critique on the
structure and dynamics of the electron is presented. The classical
electromagnetism was developed based on the macroscopic experimental laws, and
the discovery of the electron and the postulate of light quantum (or photon)
historically took place at a later time. It is by tradition that electric and
magnetic fields are associated with the electron and the photon. Could this be
the source of the foundational problems? The answer in affirmative is provided
in the present work. Similarly, it is argued that endowing rest mass (nonzero) 
to the electron is by definition; one can account for the electron dynamics
assuming it to be massless. Thus the proposed elementary objects or particles
are massless electron and neutrino, and the photon is considered a composite 
structure with neutrinos as its constituents. The fundamental entity is
postulated to be space-time bounded structure. The charge and interaction
should be explained in terms of spaced-time symmetries. Pondering over the
meaning of electronic charge, e and the fine structure constant,$\alpha  =
e^2/\hbar c = 1/137$ (where $\hbar = h/2\pi$, $h$ is the Planck's
constant, and $c$ is the velocity of light) it occurred to me that
$e^2/c$ has the dimension of the angular momentum. Does this
indicate a relationship between charge and spin or rotation? Explaining charge  
in this way also brings us closer to our goal of reducing everything to the
space-time. Here it must be mentioned that recently I discovered that the
curious dimension of $e^2/c$ was noticed by Einstein as early as  
1907 \cite{8}. This monograph thus deals with very unconventional ideas, and
suggests radical paradigm shift for the fundamental physics \cite{9}.
Photon-fluid, two dimensional space + one dimensional time physics and knot
theory are identified deserving serious
attention of the experts with a new perspective presented  here. 

\section{Speculation and experimental philosophy}

Speculation is the lifestream of the experimental science, without speculation
and hypotheses the empirical data is merely 
an information directory. Speculative ideas serve the purpose of bringing
deeper secrets on the horizon, and quite 
often stimulate meaningful experimentation and theoretical investigations. The
quote above from the Bakerian Lecture 
of Crookes \cite{5}, hypotheses proposed by Issac Newton despite his claim
`hypotheses non fingo' \cite{10} and Poincare's  influential work `La Science
et l' Hypothese' \cite{1} quite convincingly illustrate the importance of
speculation in  science. However, it is also an equally important fact that
almost always new ideas have been rejected or resisted  by the scientists who
themselves have been responsible for original work. G.P. Thomson, son of J.J.
Thomson notes \cite{11} : ``In looking back at it, one is impressed by the
extent to which a theory long held can blind even first-rate  minds to new
ideas and by how easy it is to explain almost anything in terms of a favorite
theory''. J.J. Thomson  himself recollects \cite{12}: ``At first there were
very few who believed in the existence of these bodies smaller than  atoms. I
was even told long afterwards by a distinguished physicist who had been
present at my lecture at the Royal  Institution that he thought I had been
`pulling their legs'. I was not surprised at this, as I had myself come to  
this explanation of my experiments with great reluctance $\ldots$''. The Thomsons
are referring to the controversy   regarding the nature of the cathode rays
whether they were some form of aetherial waves or material particles.   New
arguments and experiments were put forward to support one belief or the other.
The famous debate on the   interpretation of quantum mechanics between
Einstein and Bohr shows that to support one's beliefs scientists   continue
to invent newer arguments. I have attempted to understand this psychology of
scientists or rather original minds in terms of subtle transcendental
experiences which lead them to form quite often a rigid world outlook
\cite{2}. It is only unequivocal experiments which force them to dilute their
beliefs arising from their transcendental experiences. In the absence of such
experiments if the philosophical or the logical arguments are the deciding
factors they are most likely to stick to their beliefs. In the case of the
cathode rays, experiments clinched the issue in favour of material particle
interpretation, while the debate on the quantum mechanics has not ceased due
to the undecidable nature of the outcome of the most sophisticated experiments
performed till date. 

I think resistance to new ideas inspired by the philosophical beliefs based on
certain experiences is natural, unavoidable, and most of the time proves
fruitful in the quest for the knowledge. Since late 1970s, an unfortunate 
trend has gained prominence : it is not so much the scientific beliefs as the
nonscientific factors like marketism and emergence of `big science' that
unconventional simple alternatives are blocked. Do the words like `Theory of 
Everything' or the `Standard Model' reflect humility and openness expected with
the ever expanding knowledge? Superstring theory, termed as the `theory of
everything' by John Ellis \cite{13}, has remained unconnected with the 
physical world; Frank Wilczek has rightly remarked that `I don't like that
term (theory of everything), It's very, very arrogant and misleading'. Not
only this, the high energy experimental results are also presented  in such a
manner that they have acquired an aura of the ultimate knowledge. Such a faith
in a specific world-view  is certainly not good for the endeavour of
scientific truth, I believe the basic philosophy of experimental  science
demands serious rethinking on the direction and the value of contemporary
science.

\section{Wherefore high energy physics?}

The particle accelerators and the collision experiments have certainly led to
landmark discoveries, and have stimulated new and interesting physics. The
question arises : should we build the accelerators for higher and higher
energies? It would be illuminating to begin the discussion with a quotation
from an article written in 1970 by Freeman J. Dyson \cite{14}, ``there are two
main ways of doing high energy physics. The rich man's way is to build
accelerators, which give high-intensity beams of particles with accurately
controlled energy. The poor man's way is to use the cosmic rays, which descend
like the rain from heaven upon poor and rich alike, but have very low
intensity and completely uncontrolled energy. I think there is a
better-than-even chance that the major discoveries of the next 30 years in
high energy physics will be made with cosmic rays. That is why I venture to
say that it may be good for us, scientifically speaking to be poor.'' Though
Dyson's forecast proved wrong, it is worth asking: Will it be true for the
next 30 years i.e. first three decades of 21st century? I think, if we leave
aside the hope for new discoveries with cosmic rays, it is fairly reasonable
to expect that going for higher energies in laboratory will not be
scientifically productive. 

Historical importance of Rutherford scattering is well documented; in 1909 H.
Geiger and E. Marsden performed alpha ray scattering experiments \cite{15} in
Rutherford's laboratory. Thin foils of gold, approximately 0.5 micron  thick
were used as targets for the $\alpha$-particle beams which as we know today are
positively charged helium nuclei. Experimental results showed that most of the
$\alpha$-particles were deflected within an angle of 1 or 2 degree, while  
occasionally scattering at large angles of more than 45$^{\mathrm{o}}$ and
backward scattering also took place. Rutherford in 1911 proposed an atomic
model using these experiments \cite{15} in which the positively charged   
nucleus is concentrated in a radius of about 10$^{\mathrm{-12}}$ cm surrounded
by negatively charged electrons. The earlier theoretical contributions for a
planetary model of atom include those of Johnstone Stoney, J.P. Perrin and H.
Nagaoka \cite{16}. To probe deep into the structure of matter using scattering
processes has been the basic approach since then in electron-atom collisions
and elementary particle physics. I will discuss two important experiments in
high energy physics and contrast their significance : deep inelastic
electron-proton scattering (MIT-SLAC experiment) and high energy
proton-antiproton collision (CERN SPS UA1 and UA2 experiment). 

Let us first discuss the meaning of high energy in particle physics. The total
energy available for the production of new or additional particles in the
scattering process is the center of mass energy. If the four-momenta of two 
particles in the collision  are $p_{\mathrm{1}}$ and $p_{\mathrm{2}}$
(four-vector $p = (E, \bf{p}$) with masses  $m_{\mathrm{1}}$ and
$m_{\mathrm{2}}$ then the Lorentz-invariant scalar 
\begin{equation}
s  =  (p_1 + p_2)^2 = (E_1 + E_2)^2 - (\bf{p}_1 + \bf{p}_2)^2
\end{equation}
determines the center of mass energy. In the laboratory frame, let particle 2
be stationary, and the laboratory frame energy of particle 1 be $E_{1\; lab}$ 
then Eq. (1) gives 
\begin{equation}
s = m_1^2 + m_2^2 + 2E_{1\;lab}m_2
\end{equation}

In the center of mass frame, let ($E_{1cm}$, {\bf p}$_{1cm}$)
and (E$_{\mathrm{2cm}}$ , {\bf p}$_{\mathrm{2cm}}$) be the four momenta of the
particles then {\bf p}$_{\mathrm{1cm}}$ = - {\bf p}$_{\mathrm{2cm}}$ ,  and Eq.
(1) reduces to 
\begin{equation}
s=({E_{1cm}+E_{2cm}})^2
\end{equation}
In the conventional accelerators, a beam of particles is scattered from a fixed
target; Eq. (2) shows that the center of mass energy $\sqrt{s}$ increases
roughly proportional to $\sqrt{E}_{1\;lab}$ . The beam colliders employ
colliding beams of equal but opposite momenta, thus according to Eq. (3)
higher values of s can be achieved  in this case. The high energy region is
determined by the mass scale of the particles of interest or the  energy scale
of the interaction. In mid 1950s the High Energy Physics Laboratory at
Stanford utilized electron  beam energy of 0.55 GeV to study the proton
structure in the elastic scattering process  
\begin{equation}
e^- + p \to e^- + p
\end{equation}

These experiments showed that proton is not a point particle, but an extended
structure. Compare the high energy region four decades later for the discovery
of the sixth quark, top t. The top quark mass is a free parameter in the SM,
and is believed to be $\sim 200$ times the mass of proton though the reason for
this is not known. The $e^+\; e^-$ accelerators at CERN
and SLAC operate at $\sqrt{s}\sim 91$ GeV,  but $M_t$ > 45
GeV/c$^2$ therefore the top quark cannot be observed in these accelerators.  The top
quark was finally discovered in 1994 \cite{3} at the Fermilab Tevatron $p{\bar p}$
collider with $\sqrt{s}$ = 1800 GeV = 1.8 TeV;  the measured mass
$M_t = 175 \pm$ 8 GeV/c$^2$. The new $p{\bar p}$ collider LHC at CERN  with 
$\sqrt{s}$ = 14 TeV is expected to be operational within next few years.

\subsection{The MIT-SLAC Experiment}

The deep inelastic electron-proton scattering experiment has
the same significance for the nucleons as the  Geiger-Marsden-Rutherford
$\alpha$-particle scattering had for the atomic structure \cite{15}. Inelastic
scattering of electrons  with liquid hydrogen and liquid deuterium targets was
started in 1967 as MIT-SLAC collaborative project using  electron beams with
the highest energy of $\sim 21$ GeV in an underground two miles long
accelerator. An idea of  the tremendous amount of ingenuity, dedication and
team-work required for building the machine and the high  energy spectrometer
can be had from the account given by Taylor \cite{6}. For detailed theoretical
treatment, we  refer to a monograph exclusively on the deep inelastic
scattering \cite{17}, and also \cite{6}. Inelastic scattering is the  process
\begin{equation}
e^- + p \to e^- + X
\end{equation}
where $X$ denotes one or more hadrons in the final state. This is an example of
an inclusive reaction in which only the scattered electron is detected. For the
collision process (5) one can define the following kinematic variables
\begin{eqnarray}
q^2 &=& ({p-p\prime})^2 \\
\nu &=& \frac{p.q}{M} \\
M_X^2 &=& ({p+q})^2  
\end{eqnarray}
Here $p$ and $p^\prime$  are the 4-momenta of the incoming and scattered electron, $P$
and $M$ are the 4-momentum and mass of the proton respectively, $q$, is the
4-momentum of the virtual photon, and $M_X$ is the total mass of
the  hadron(s) $X$. In the laboratory frame, neglecting the mass of the
electron, we have  
\begin{eqnarray}
q^2 &=& -Q^2 = -4E E' \sin^2\frac{\theta}{2} \\
\nu &=& E - E'   \\
M_X^2 &=& M^2 - Q^2 + 2M\nu
\end{eqnarray}
where $E$, $E'$ are the energies of the incident and scattered electron
respectively and $\theta$ is the scattering angle.  A dimensionless variable $x$
is often used, defined by  
\begin{equation}
x = \frac{Q^2}{2M\nu}
\end{equation}
It can be shown that the differential cross section for the process (5) may
be calculated using the Feyman diagram for this reaction to the lowest order
electromagnetic electron-proton-coupling via the exchange of a virtual photon,
and expressed in terms of the structure functions $W_1$ and
$W_2$ as follows 
\begin{equation}
\frac{d^2\sigma}{d\Omega dE'} = \sigma_M [W_2 + 2W_1 \tan^2\frac{\theta}{2}]
\end{equation}
Here $\sigma_M$ is the Mott cross section
\begin{equation}
\sigma_M = \frac{\alpha^2 \cos^2 \theta/2}{4E^2 \sin^4 \theta/2}
\end{equation}

Both $W_1$ and $W_2$ are functions of momentum transfer, $Q^2$ and energy loss, $\nu$. 
In order to determine the structure functions, the differential cross section at 
several values of the angle $\theta$ for fixed $\nu$ and $Q^2$ has to be measured. The early 
experimental data  showed two  unexpected features: (1) the inelastic cross section was 
found to have weak dependence on the momentum transfer for large $Q^2$, and
(2) in the asymptotic region where both $\nu$ and $Q^2$ become very large $\to \; \infty$ 
keeping $x$ to be finite, the structure functions showed the scaling behaviour
\begin{eqnarray}
\nu W_2 (Q^2, \nu) & = & F_2(x) \\
2 M W_1(Q^2, \nu) & = & F_1 (x) 
\end{eqnarray}

In view of the existing contemporary interpretation of the elastic scattering
data, both results appeared rather puzzling. The cross section for the process
(4) was known to fall rapidly with the increasing momentum transfer as
compared to that from a point charge. The electric and magnetic form factors
in the cross section satisfied the dipole form for $Q^2 \leq 10$ GeV $^2$.
\begin{equation}
G(Q)^2 = \left(1 + \frac{Q^2}{0.71{\rm GeV^2}}\right)^{-2}
\end{equation}
The accepted interpretation of these results was that the proton is not a point
particle, but has a diffused extended structure with a size of $\sim 0.8$ Fermi.
The deep (large $Q^2$) or continuum inelastic scattering experiments, 
on the other hand, seemed to indicate electron scattering from
point particles. The question arose whether proton had internal structure with
more elementary constituents. If yes, what are they?  Though already quark
model for hadron spectrum was there, it was thought more like a book keeping
framework than a possible dynamical theory for strong interaction at that
time. Bjorken's conjecture for scaling behavior using current algebra sum
rules in 1967 prior to the MIT-SLAC data did not attract immediate attention. 
It has been point out \cite{6} that Feynman's interpretation of the data in terms of
the parton model in 1969 gave impetus to the model of the internal structure of
the nucleons. For a critical evaluation of the parton model in the deep
inelastic process, see \cite{18}. Assuming partons to be point particles, and to be 
noninteracting with each other during virtual photon absorption one can
calculate the cross section for the process (5). The variable $x$ defined by
Eq. (12) turns out to correspond to the ratio of the parton's momentum to
the proton's momentum i.e. if $N$ partons constitute a proton, then the momentum  
$P_i$ of the $i^{th}$ parton in the infinite momentum frame is given by
$P_i$ = $x_i P$, where $x_i$ lies between 0 and 1. The infinite momentum frame 
is another way of looking at free partons. Let us assume that in the rest frame 
partons interact with each other changing their momenta during finite time 
intervals. As the momentum increases, the
Lorentz transformation to the infinite momentum frame gives time dilation such
that the changes occur so slow that the partons appear to be free. In this
approximation scattering from a point charge e$_{\mathrm{i}}$ of a parton can
be calculated to give 
\begin{equation}
W_2^i(Q^2,\nu) = e_i^2 \delta\left(\nu -\frac{Q^2}{2Mx_i}\right)
\end{equation}
and the structure function
\begin{equation}
W_2(Q^2,\nu) = \sum_i \int_0^1 f_i(x_i) W_2^i dx_i
\end{equation}
Here $f_i(x_i)$ is the probability of finding the $i^{th}$ parton with the 
momentum fraction $x_i$. Substituting $W_2^i$ from Eq. (18) and carrying  our the
integration in Eq. (19) finally we get 
\begin{equation}
\nu W_2(Q^2, \nu) = \sum_i e_i^2f_i(x)x
\end{equation}
Comparing this equation with (15), we recognize that parton picture leads to
Bjorken scaling. The calculation of scattering amplitude for scalar and spinor
parton shows that while $W_2$ is unchanged, $W_1$ is zero for the former. 
In fact, for spin $1/2$ partons 
\begin{equation}
F_2(x) = xF_1(x)
\end{equation}
and for spin zero partons
\begin{equation}
F_1(x) = 0
\end{equation}
in the Bjorken limit. Experiments show the behaviour expressed by Eq. (21),
therefore, it can be concluded that partons are spin $1/2$ particles. Are they
quarks? This identification is not straightforward, however, a dynamical 
theory of quark-quark interaction with gluons as strong gauge fields, namely,
the quantum chromodynamics was soon developed. The prediction of the
logarithmic deviations from the Bjorken scaling  confirmed in muon and 
neutrino scattering from nucleons gave confidence in the QCD. The theory was
shown to be asymptotically free, which explained the assumption of free
partons mentioned earlier. 

\subsection{The discovery of the weak gauge bosons}

The discovery of the weak gauge bosons in the Super Proton Synchrotron $p{\bar p}$
collider at CERN is considered a great milestone in the quest for unified
theory. The production of W, Z with masses about 80 and 90 times the proton
mass respectively was expected in the proton-antiproton collision at $\sqrt{s}$ = 540
GeV achieved in the SPS collider. The idea of stochastic cooling pioneered by
S. van der Meer, and sophisticated advanced electronics made it possible to
accelerate and accumulate the $p{\bar p}$ beams to such a high energy range. In the SPS
ring, 2.2 Km in diameter, the proton and artiproton beams accelerated to 26
GeV/c in the PS machine are injected in the opposite directions, and
accelerated to high energy of 270 GeV. They are bunched for collisions to take
place at well defined locations in the SPS ring. Antiprotons are created
bombarding 26 GeV/c protons on Cu target in the PS. An accumulator ring in one
day accumulates about 10$^{11}$ antiprotons, which are accelerated to 26 GeV/c in the PS.

In the generation of intense antiproton beams the stochastic cooling has a key
role. Randon motion of particles in the beam is observed by pick-up sensors,
and the signal is used in a kicker to push the particles towards a desired
position \cite{19}. Since the spread of the momenta of the particles is reduced in
this process, it is referred to as beam cooling. The stochastic cooling is
used in the antiproton storage ring, the Antiproton Accumulator of the SPS.
The first project on $p{\bar p}$ collision was code-named UA1 (\uline{U}nderground
\uline{A}rea), and was led by Carlo Rubbia, and the second experiment UA2 was 
led by Pierre Darriulat.

In the search for Ws, the reaction is 
\begin{equation}
p + {\overline p} \to W^{\pm} + X
\end{equation}
and from the decay mode 
\begin{equation}
W^{\pm} \to e^{\pm} + \nu_e
\end{equation}
detection of electrons and missing energy in the form of neutrinos provide
hints for the Ws. Here X denotes other particles `the sum of the debris from
the interactions of the other protons' \cite{19}. The process
\begin{equation}
Z \to e^+ + e^- \qquad {\rm or} \qquad \mu^+ + \mu^-
\end{equation}
is a factor of 10 less probable than (23), however, the leptonic decay modes
are easier to detect.

In January, 1983 six possible W events by UA1 and four by UA2 were announced;
observing high energy electrons in the detectors looking for them at
relatively large angles to the beam direction. High energy particle tracks in
opposite directions as a signature for Z neutral gauge boson were observed in
both UA1 and UA2 detectors. The discovery was announced in June, 1983 based on
2 or 4 Z events \cite{19}. Since then Fermi lab. in 1985, and the Stanford Linear
Collider later detected the weak gauge boson events in large numbers.

\subsection{Contrasting the two : alternatives}

The first important difference in the two experiments is regarding
the motivation. The MIT-SLAC experiment was planned to study the
electro-production of resonances as a function of momentum transfer, and to
probe the inelastic continuum in the high energy region. The unexpected
results led to the discovery of the internal constituent model of the
nucleons. On the other hand, in the SPS collider, the experiment was set to
see the  W and Z events almost with certainty. 
Equally important point distinguishing the two is the role of skepticism in
analyzing the data. The first hint for the W events, few in number in millions
of collisions, came in the beginning of January, 1983; and on 25 January, 1983
the discovery was announced in a Press Conference at CERN, more like a
dramatic event.  In contrast to this, the deep inelastic scattering
experiments were carried out with thorough analysis. To quote from Kendall
[6], ``The collaboration was aware from the outset of the program that there
were no accelerators in operation, or planned, that would be able to confirm
the entire range of the results.  The group carried out independent data
analyses at MIT and at SLAC to minimize the chance of error. One consequence
of the absence of comparable scattering facilities was that the collaboration
was never pressed to conclude either data taking or analysis in competitive
circumstances. It was possible throughout the program to take the time
necessary to complete work thoroughly''. 

Finally, the approach for theoretical interpretation is markedly different. The
weak gauge bosons' signatures were immediately identified confirming the SM.
The results from the MIT-SLAC experiment led to intense debate comparing the
parton model with other competing non-constituent models like the
vector-dominance model and Regge exchange mechanisms.

To conclude, more than the discoveries of tau-lepton in 1975 and b-quark ($b{\bar b}$
bound state) in 1977, it is the discovery of W and Z that gave big boost for
building the accelerators at still higher energies. Most of the particle
physicists believe that TeV energy range is absolutely essential for new
physics.  Such an approach seems unsatisfactory for at least two reasons: the
maximum energy in the laboratory with best possible resources and funding is
unlikely to reach 10$^{\mathrm{15}}$ GeV  (the Grand Unification Scale), and
secondly the indirect rare signatures in the TeV range will be extremely
difficult to interpret, more so in view of many speculative models in between
the SM and the GUT. Thus the need for alternative strategy is forced on us for
down to earth practical reasons. 

A logical approach in the best of the scientific traditions is to do precision
experiments using the existing facilities. This program has already started
\cite{20}, and deserves more  attention and importance. It would be less expensive,
and has a potential to probe new physics, if any. Exploring low energy 
physics afresh in the light of rich empirical data obtained in high energy
experiments may also prove fruitful. For example, study of protonium ($p{\bar p}$ bound
state) spectroscopy seems feasible in view of the recent remarkable success in
creating anti-hydrogen (${\bar p}e^+$ bound state) at CERN. Interesting results  on the
strong force may be expected from this. Low energy scattering for quark
ionization such that scattered particle becomes fractionally charged seems
another possible idea. Innovative ideas in this direction need to be
encouraged.

In the present work, rethinking on the entire approach towards unification and
ultimate reality is suggested: the principle of simplicity and parsimony guides
us for searching the alternative. Would it not be the simplest idea if the
space-time is the fundamental physical entity? Without postulating any new
elementary constituents, the proposition that electron and neutrino are the 
elementary constituents of matter, is put forward to stimulate further
investigations, and revision of the current focus on high energy physics.

\section{Plan of the book}

I have explained that this book is written with a radically new viewpoint on
the foundational problems, however, technical rigour and scientific accuracy of 
the subject matter under discussion have been maintained. The reader 
with a background in field theory (both classical and quantum) should be able
to appreciate the arguments presented.  Quotations from the original writings are 
used to convey significant and unorthodox views of the writers, and 
sometimes just because I found them exceptionally lucid and effective.
Exhaustive and complete review citing the work of all active researchers is not 
claimed, but the important contributions relevant for our arguments
 have been included. The Source Book by Magie and Whittaker's two volumes are
referred to in the text by (M page number) and W (W vol page number) respectively due 
to frequent citations.

The organization of the text is such that one may classify it into four categories: 
\begin{enumerate}
\item
second Chapter reviews the Standard Model for unified strong, weak and electromagnetic 
forces with a critical commentary, 
\item
next three Chapters are devoted to the physical properties of the electron, neutrino 
and photon; their present understanding, outstanding problems and alternative ideas,
\item
Chapters six to eight deal with the classical electrodynamics with emphasis on the 
foundational problems, attempts for the modifications, and field theory in the Weyl space,
and 
\item
the last Chapter propounds a tentative model of the electron and outlines significance
of three (2+1) dimensional field theories and knot theory for building an alternative model
for the elementary particles and their interactions.
\end{enumerate}


\begin{thebibliography}{99}
\bibitem{1}
H. Poincare, The Foundations of Science (The Science Press,1946)
\bibitem{2}
S. C. Tiwari, Phys. Essays 7, 22, 1994; in Proceedings Conference on Physical
Interpretation of Relativity Theory(PIRT), British Society for the Philosophy
of Science, London (September 1992 pp375-386)
\bibitem{3}
Review of Particle Physics, Phys. Rev. D54 Part-I,1996
\bibitem{4}
T. D. Lee, CERN Courier 27,7,1987
\bibitem{5}
W. Crookes, Phil. Trans., Part-I,135,1879
\bibitem{6}
R.E. Taylor, H.W. Kendall and J.I. Friedman, Rev. Mod. Phys.63,573-629,1991
\bibitem{7}
P.A.M. Dirac,Directions in Physics (John Wiley 1978)
\bibitem{8}
H.Woolf, ed. Some Strangeness in the Proportion (Addison-Wesley 1980)
\bibitem{9}
S. C. Tiwari, in Global Conference on Mathematical Physics (Einstein
Foundation International Nagpur,India 1987)
\bibitem{10}
S.C. Tiwari, Phys. Essays 2,313,1989
\bibitem{11}
G.P. Thomson, Phys. Today 20(5),55,1967
\bibitem{12}
J.J. Thomson, Recollections and Reflections (Bell \& Sons Ltd. London 1936)
\bibitem{13}
F. Flam, Science 256,1518,1992
\bibitem{14}
F. J. Dyson, Phys. Today, 23(9),23,1970
\bibitem{15}
H. Geiger and E. Marsden, Proc. Roy. Soc. (London) 82,495,1909; E.
Rutherford, Phil. Mag. 21,669,1911
\bibitem{16}
D. L. Anderson, The Discovery of the Electron (D. van Nostrand Co. Inc. 1964)
\bibitem{17}
R.G. Roberts, The Structure of the Proton (C.U.P. 1990)
\bibitem{18}
R.P. Feynman, Photon-Hadron Interactions (Reading, Mass, 1972)
\bibitem{19}
S. van der Meer, C. Rubbia , Rev. Mod. Phys. 57.689-722.1985
\bibitem{20}
P. Langacker, M.Luo, A.K. Mann, Rev. Mod. Phys. 64,87,1992
\end{thebibliography}
\end{document}